\PassOptionsToPackage{english}{babel}
\documentclass[a4paper,twocolumn,11pt,accepted=2022-02-10]{quantumarticle}
\pdfoutput=1
\usepackage[T1]{fontenc}
\usepackage{hyperref}

\usepackage[english]{babel}
\usepackage{graphicx}  
\usepackage{dcolumn}   
\usepackage{bm}        
\usepackage{bbm}        
\usepackage{amssymb}   
\usepackage{amsmath}
\usepackage{mathtools}
\usepackage[utf8]{inputenc}
\usepackage{siunitx}

\usepackage{changes}

\definecolor{myurlcolor}{rgb}{0,0,0.7}
\definecolor{myrefcolor}{rgb}{0.8,0,0}
\usepackage{hyperref}
\hypersetup{colorlinks, linkcolor=myrefcolor,
citecolor=myurlcolor, urlcolor=myurlcolor}

\usepackage{cleveref}
\crefrangeformat{equation}{#3(#1 #4--#5 #2)#6}

\usepackage{comment}
\usepackage{soul} 
\newcommand{\ignore}[1]{}
\usepackage{enumitem}

\newcommand{\eg}{\textit{e.g. }}				
\newcommand{\ie}{\textit{i.e. }}				

\newcommand{\bra}[1]{\ensuremath{\langle#1|}}
\newcommand{\ket}[1]{\ensuremath{|#1\rangle}}

\newcommand{\abs}[1]{\ensuremath{\left\vert #1\right\vert}}

\newcommand{\avg}[1]{\langle#1\rangle}		
\newcommand{\var}{\text{Var}}		

\newcommand{\mc}[1]{\mathcal{#1}}

\newcommand{\fisher}{F}

\newcommand{\qfi}{\fisher_Q}
\newcommand{\cqfi}{\fisher_Q^{B|A}}

\newcommand{\assem}{\mc{A}}

\newcommand{\ox}{\otimes}	
\newcommand{\id}{\mathbbm{1}}

\begin{document}
\selectlanguage{english}

\title{Entanglement of Local Hidden States}
\author{Matteo Fadel}
\email{fadelm@phys.ethz.ch} 
\affiliation{Department of Physics, ETH Z\"urich, 8093 Z\"urich, Switzerland} 
\affiliation{Department of Physics, University of Basel, Klingelbergstrasse 82, 4056 Basel, Switzerland}
\orcid{0000-0003-3653-0030}

\author{Manuel Gessner}
\email{manuel.gessner@icfo.eu}
\affiliation{ICFO-Institut de Ci\`{e}ncies Fot\`{o}niques, The Barcelona Institute of Science and Technology, Av. Carl Friedrich Gauss 3, 08860, Castelldefels (Barcelona), Spain}
\affiliation{Laboratoire Kastler Brossel, ENS-Universit\'{e} PSL, CNRS, Sorbonne Universit\'{e}, Coll\`{e}ge de France, 24 Rue Lhomond, 75005, Paris, France}
\orcid{0000-0003-4203-0366}

\date{\today}

\begin{abstract}
Steering criteria are conditions whose violation excludes the possibility of describing the observed measurement statistics with local hidden state (LHS) models. When the available data do not allow to exclude arbitrary LHS models, it may still be possible to exclude LHS models with a specific separability structure. Here, we derive experimentally feasible criteria that put quantitative bounds on the multipartite entanglement of LHS. Our results reveal that separable states may contain hidden entanglement that can be unlocked by measurements on another system, even if no steering between the two systems is possible.
\end{abstract}

\maketitle

\section{Introduction} The classification of quantum correlations is crucial for our understanding of the resources that enable quantum information tasks~\cite{NielsenChuang, HorodeckiRMP, ReidRMP2009, BrunnerRMP, CavalcantiREPPROGPHYS17, FriisNATREV2019, GuehneRMP2020}. Two of the main challenges in this field are the characterization of entanglement in multipartite systems~\cite{DurPRA2000,GuhneToth,LeviPRL2013,WalterArXiv,SzalayQUANTUM2019} and the identification of nonclassical correlations stronger than entanglement that allow to relax the assumptions to be made about the system of interest~\cite{ReidRMP2009, BrunnerRMP, CavalcantiREPPROGPHYS17, GuehneRMP2020}. Bell nonlocality, the strongest form of quantum correlations, allow for fully device-independent entanglement detection~\cite{Bell1964,BrunnerRMP}. Einstein-Podolsky-Rosen (EPR) steering represents an intermediate notion between entanglement and nonlocality~\cite{WisemanPRL2007}, and it allows for one-sided device-independent entanglement~\cite{GallegoPRX2015,GuehneRMP2020}.

The observation of an EPR paradox~\cite{EPR1935}, or more generally of steering, formally excludes the possibility to describe the observed data in terms of a local hidden state (LHS) model that assigns local quantum states to one of the subsystems~\cite{WisemanPRL2007}. Steering criteria rule out LHS descriptions by verifying a local complementarity principle in form of an uncertainty relation~\cite{ReidPRA1989} or a metrological bound~\cite{YadinNC2021}. Uncertainty-based criteria formulated in terms of variances are most widely used in experiments and are particularly suited to reveal the steering of approximately Gaussian states~\cite{ReidRMP2009}. Metrological approaches~\cite{YadinNC2021,GuoArXiv}, as well as entropic relations~\cite{WalbornPRL11,WalbornPRA13,CostaEntropy2018,RiccardiPRA2018}, have particular advantages for detecting steering of non-Gaussian states. So far, despite advances in multipartite steering~\cite{HePRL2013,ReidPRA2013,TehPRA2014,CavalcantiNatCommun2015,ArmstrongNatPhys2015,CostaEntropy2018,RiccardiPRA2018,XiangPRA2019}, these criteria are mostly applied in bipartite settings. Because of the weaker assumptions that can be made about the system, detecting steering is generally challenging and requires a higher degree of control than entanglement detection.

In this work we show that even if the recorded data is unable to rule out all LHS descriptions, it may still pose limitations on the classes of LHS that are capable of reproducing the observed correlations. We focus on multipartite settings with one untrusted party ($A$) that may share quantum correlations with another multipartite quantum system ($B$), on which LHS models can be classified according to their multipartite entanglement. We derive families of criteria that exclude LHS models with specific separability properties. In the limit when even genuine multipartite entangled LHS are excluded these criteria converge to metrological steering criteria that can be approximated to yield uncertainty-based bounds. By resolving the substructure of LHS models our results provide further insight into the manifestations of nonclassical correlations in multipartite systems. For instance, we show that if a LHS model for $B$ exists, this model cannot always be constructed from separable LHS, even if $B$ is separable. This implies that quantum information processing assisted by measurements on $A$ can unlock hidden entanglement of $B$ even in the absence of steering. Our criteria can be tested with state-of-the-art experimental setups and provide quantitative bounds on the multipartite entanglement of LHS models.

\section{Separable LHS models}
We consider a multipartite setting with one untrusted party $A$ (Alice), and a multipartite quantum system $B$ (Bob). The joint measurement statistics can be modeled in terms of the assemblage $\mathcal{A}(a,X)=p(a|X)\rho_{a|X}^B$, where $a$ is the result of measuring $X$ on $A$, and $\rho_{a|X}^B$ are conditional quantum states for $B$ \cite{CavalcantiREPPROGPHYS17}. EPR steering from $A$ to $B$ is observed as an assemblage which fails a description in terms of a LHS model~\cite{WisemanPRL2007} of the form $\assem(a,X) = \sum_\lambda p(\lambda) p(a|X,\lambda) \sigma^B_\lambda$, where $p(\lambda)$ is a probability distribution for a local hidden variable $\lambda$ that determines both Alice's local probability distribution $ p(a|X,\lambda)$ and Bob's LHS $\sigma^B_\lambda$.

We define separable LHS models as assemblages whose LHS are subject to additional separability constraints. We first focus on separability in a specific multipartition, and then discuss bounds that exclude the convex hull of families of partitions. Let $\Lambda=\{\mathcal{B}_1,\dots,\mathcal{B}_{|\Lambda|}\}$ be a partition of $B$ into $|\Lambda|$ subsets $\mathcal{B}_k$, each containing $|\mathcal{B}_k|$ parties such that $\sum_{k=1}^{|\Lambda|}|\mathcal{B}_k|=N$. A state $\sigma_{\Lambda-\mathrm{sep}}$ is separable in the partition $\Lambda$ if there exist local quantum states $\sigma_\gamma^{\mathcal{B}_k}$ and a probability distribution $p(\gamma)$ such that $\sigma_{\Lambda-\mathrm{sep}}=\sum_{\gamma}p(\gamma) \sigma^{\mathcal{B}_1}_{\gamma}\otimes \cdots \otimes \sigma^{\mathcal{B}_{|\Lambda|}}_{\gamma}$. Accordingly, a $\Lambda$-separable LHS model is described by a LHS model where all LHS $\sigma^B_\lambda$ are chosen $\Lambda$-separable.

\section{Metrological detection of inseparable LHS}
In order to witness the separability structure of LHS models, we derive bounds on the average metrological sensitivity of Bob. To this end, we consider measurements of a phase shift $\theta$ generated by the Hamiltonian $H=\sum_{i=1}^NH_i$, where each $H_i$ acts locally on Bob's subsystem $B_i$. Without information from Alice, Bob's system is described by the reduced density matrix $\rho^B=\sum_ap(a|X)\rho^B_{a|X}$. By choosing an optimal measurement observable, he is able to extract the full metrological sensitivity of the state $\rho^B$, which is described by the quantum Fisher information (QFI) $F_Q[\rho^B,H]$~\cite{BraunsteinPRL1994,GiovannettiNatPhoton2011,ApellanizJPA2014,Varenna}. An upper bound for this sensitivity is given by $F_Q[\rho^B,H]\leq 4\var[\rho^B,H]$, which describes a complementarity between Bob's phase sensitivity and the fluctuations for measurements of the generator $H$~\cite{YadinNC2021}.

Assume now that Alice performs a measurement $X$ and obtains the result $a$. This will project Bob's system into the conditional state $\rho^B_{a|X}$. If the information $(a,X)$ is provided to Bob, he can adapt the choice of his measurement observable such that it optimally extracts the sensitivity of the conditional state, leading to the sensitivity $F_Q[\rho^B_{a|X},H]$. Since Alice's results occur randomly, Bob's average sensitivity is given, after an optimization over Alice's setting $X$, by the quantum conditional Fisher information (QCFI)~\cite{YadinNC2021}
\begin{align}\label{eq:assistedFQ}
F_Q^{B|A}[\mathcal{A},H]:=\max_{X}\sum_{a}p(a|X)F_Q[\rho^B_{a|X},H].
\end{align}
Similarly, using another measurement setting, Alice may remotely prepare conditional states for Bob that have small variances for measurements of the generator $H$ and yield the quantum conditional variance (QCV)~\cite{ReidRMP2009,YadinNC2021}
\begin{align}
\var_Q^{B|A}[\mathcal{A},H]:=\min_{X}\sum_{a}p(a|X)\var[\rho^B_{a|X},H].
\end{align}

Let us now assume that the correlations in the multipartite system can be described by a separable LHS model. One of our main results is the upper bound on the QCV for $\Lambda$-separable LHS models (see the Supplemental Material~\cite{Supp} for details)
\begin{align}\label{eq:mainLambda}
    \qfi^{B_1\dots B_N \vert A}[\assem,H]   &\leq 4\sum_{k=1}^{|\Lambda|}\var_Q^{\mathcal{B}_k|A}[\assem,H^{\mathcal{B}_k}].
\end{align}
Here, $\var_Q^{\mathcal{B}_k|A}[\assem,H^{\mathcal{B}_k}]$ describes the QCV for subsystem $\mathcal{B}_k$. A violation of condition~(\ref{eq:mainLambda}) hence implies that the observed sensitivity and fluctuations cannot be described in terms of a $\Lambda$-separable LHS model, either because steering from $A$ to $B$ is possible, or because only entangled LHS models are able to account for the correlations.

\section{Relation with metrological steering and entanglement criteria}
The existence of a separable LHS model implies (i) that no steering from Alice to Bob is possible and (ii) that Bob's system is separable. It is interesting to observe how independent criteria for (i) and (ii) can be derived from~(\ref{eq:mainLambda}). First, note that dropping all assumptions about Bob's separability properties corresponds to allowing for $\Lambda$-separable models where $\Lambda=\{B_1\dots B_N\}$ is the trivial partition with only a single system that contains all of Bob's subsystems. In this case, Eq.~(\ref{eq:mainLambda}) reduces to the metrological steering bound $\qfi^{B_1\dots B_N \vert A}[\assem,H] \leq 4\var_Q^{B_1\dots B_N|A}[\assem,H]$ which indeed holds for arbitrary LHS models~\cite{YadinNC2021}.

Second, we show that~(\ref{eq:mainLambda}) implies a known condition for $\Lambda$-separability of Bob. To this end, we note that convexity of the QFI and concavity of the variance imply that assisted measurements always perform on average at least as good as local measurements without assistance~\cite{YadinNC2021}. More precisely, we obtain $F_Q^{B|A}[\mathcal{A},H]\geq F_Q[\rho^{B},H]$ and $\var_Q^{\mathcal{B}_k|A}[\mathcal{A},H^{\mathcal{B}_k}]\leq \var[\rho^{\mathcal{B}_k},H^{\mathcal{B}_k}]$, where $\rho^{\mathcal{B}_k}$ is the reduced density matrix of subsystem $\mathcal{B}_k$. Inserting these bounds into~(\ref{eq:mainLambda}) yields $F_Q[\rho^{B},H]\leq 4\sum_{k=1}^{|\Lambda|}\var[\rho^{\mathcal{B}_k},H^{\mathcal{B}_k}]$, a necessary condition for the $\Lambda$-separability of Bob~\cite{GessnerPRA2016}. This also implies that finding a separable LHS model is impossible if Bob's reduced state is already entangled.

\section{Separable states with inseparable LHS models}
One may wonder whether these conclusions also hold in the reverse direction, \ie whether a separable LHS model can be constructed in the absence of steering if additionally Bob's system is separable. However, a simple counter-example illustrates that even the combination of both conditions does not imply the existence of a separable LHS model. Let $\ket{\uparrow}$ ($\ket{\downarrow}$) be the eigenstate of the $\sigma_z$ operator with eigenvalue $+1$ ($-1$), and let $\ket{\Phi_\pm}_B = ({\ket{\downarrow\downarrow}_B} \pm\ket{\uparrow\uparrow}_B)/\sqrt{2}$ denote Bell states for system $B=B_1B_2$. Consider the state $\rho^{AB} = \frac{1}{2}( \ket{\uparrow}\bra{\uparrow}_A \otimes \ket{\Phi_+}\bra{\Phi_+}_B + \ket{\downarrow}\bra{\downarrow}_A\otimes \ket{\Phi_-}\bra{\Phi_-}_B)$. Clearly, a LHS for $B$ exists and Bob's reduced state $\rho^B=\frac{1}{2}\left( \ket{\downarrow\downarrow}\bra{\downarrow\downarrow} + \ket{\uparrow\uparrow}\bra{\uparrow\uparrow} \right)$ is separable.
Note that every bipartite separable state can be interpreted as a mixture of entangled states~\cite{Cohen98,Cohen99}.

To test the separability of all possible LHS models that describe the correlations of this state, we make use of the condition~(\ref{eq:mainLambda}) for the bipartition $\Lambda=\{B_1,B_2\}$ and the collective spin Hamiltonian $H=J_z=(\sigma_z^{B_1}+\sigma_z^{B_2})/2$. To determine the left-hand side, we consider measurements of Alice of $X=\sigma_z$, which leads to the lower bound $F_Q^{B|A}[\mathcal{A},J_z]\geq (F_Q[\Phi_+,J_z]+F_Q[\Phi_-,J_z])/2=4$ where $\mathcal{A}$ describes the assemblage that is obtained from the bipartite state $\rho^{AB}$. This bound is tight because of the generally valid upper bound $F_Q^{B|A}[\mathcal{A},J_z]\leq 4 \var[\rho^{B},J_z]$~\cite{YadinNC2021} with $\var[\rho^{B},J_z]=1$. It is easy to see that the measurement of any spin direction by Alice projects the two subsystems $B_1$ and $B_2$ into maximally mixed states 
and we obtain $\var_Q^{B_i|A}[\mathcal{A},\sigma_z^{B_i}/2]=1/4$ for $i=1,2$. 

\begin{figure*}[tb]
	\begin{center}
		\includegraphics[width=\textwidth]{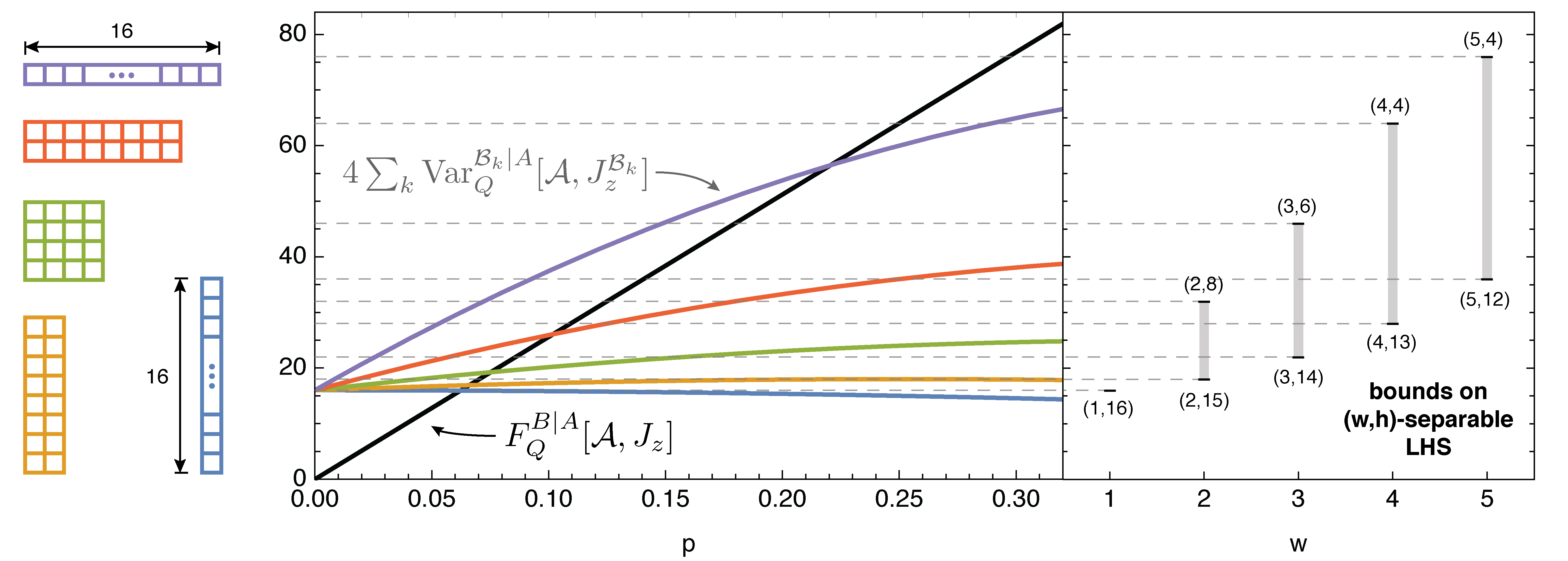}
	\end{center}
	\caption{\textbf{Inseparable LHS of noisy GHZ states.} For GHZ states mixed with white noise with probability $1-p$, $\Lambda$-separable LHS models are ruled out, according to~(\ref{eq:mainLambda}), when the QCFI (the black line represents a lower bound) surpasses the average QCV (colored lines represent upper bounds). Different partitions $\Lambda$ that separate Bob's qubits into entangled groups of equal size $N_k=N/k$ are illustrated by Young diagrams on the left, ranging from fully separable (blue) to genuine multipartite entangled (violet), the latter representing the steering bound. The right panel show state-independent bounds on the QCFI for $(w,h)$-separable LHS with different values of $w$ (see text).}
	\label{fig:exGHZ}
\end{figure*}

The violation of the condition~(\ref{eq:mainLambda}) implies that a separable LHS model for the state $\rho^{AB}$ does not exist, despite the existence of a general LHS model and the separability of Bob's reduced state. The entanglement of the LHS is generally not available to Bob, whose system appears separable under all measurements that are carried out locally. The fact that the LHS must be entangled now allows Alice to remotely prepare Bob's system in an entangled state, conditioned on the measurement result she observes. If this result is communicated to Bob, he can exploit the entanglement of his conditional state for the implementation of quantum information tasks.

\section{Noisy GHZ state}
To further illustrate these ideas, consider a system composed of $N+1$ qubits, partitioned into a single control qubit (Alice) and the remaining $N$ qubits on Bob's side. The system is prepared in a noisy Greenberger-Horne-Zeilinger (GHZ) state $\rho^{AB} = p\ket{\mathrm{GHZ}_\phi^{N+1}}\bra{\mathrm{GHZ}_\phi^{N+1}} + (1-p)\id/2^{N+1}$, with $\ket{\mathrm{GHZ}_\phi^{N+1}} = ( \ket{\downarrow}  {\ket{\downarrow}}^{\ox N} + e^{i\phi} \ket{\uparrow}  {\ket{\uparrow}}^{\ox N}) / \sqrt{2} $ \cite{GHZ}. Note that Bob's subsystem is always separable, independently of $p$. For the assemblage $\assem$ corresponding to the state $\rho^{AB}$, a lower bound to the QCFI is obtained by considering that Alice measures in the $\sigma_z$-basis and reads $\cqfi[\assem,J_z] \geq p^2 N^2/[p+2(1-p)/2^N]$~\cite{YadinNC2021}. From measurements in the $\sigma_x$ basis, we obtain the upper bound on the QCV of one of Bob's subsystems as $4\var^{\mathcal{B}_k|A}_Q[\assem,J_z^{\mathcal{B}_k}] \leq (1-p)N_k(1 + pN_k)$, where $N_k$ is the number of qubits contained in $\mathcal{B}_k$. Considering partitions of Bob into $k$ subsystems of equal size $N_k=N/k$, we obtain in the limit of large $N$ the condition $p\gtrsim k/(N(k-1))$ for inseparability of the LHS, whereas steering is only detected for $p\gtrsim 1/\sqrt{N}$. In Fig.~\ref{fig:exGHZ} we illustrate the LHS separability bounds for different subsystem sizes for a system of $N=16+1$ qubits.

\section{Variance-based criteria}
For practical implementations it is convenient to formulate witnesses involving low-order moments, in particular variances~\cite{HofmannPRA2003,GuhneToth,ReidRMP2009,TehPRA2014,FriisNATREV2019,FadelPRA2020}. Our metrological condition for separable LHS implies a weaker variance-based criterion that is close in spirit to Reid's seminal condition for arbitrary LHS models~\cite{ReidPRA1989,ReidRMP2009}. The QCFI can be approximated from below in terms of first and second moments as $\abs{\avg{[H,M]}_{\rho^{B}}}^2/\var_Q^{B|A}[\assem,M]\leq\qfi^{B \vert A}[\assem,H]$~\cite{YadinNC2021}, where $M$ is an arbitrary observable. This approximation converts the condition~(\ref{eq:mainLambda}) into the variance condition
\begin{align}\label{eq:ReidSepLHS}
&\left(\sum_{k=1}^{|\Lambda|} \var_Q^{\mathcal{B}_k|A}[\assem,H^{\mathcal{B}_k}]\right)\var_Q^{B|A}[\assem,M] \notag\\&\qquad\geq \frac{\abs{\avg{[H,M]}_{\rho^{B}}}^2}{4}.
\end{align}
In the special case of the trivial partition $\Lambda=\{B_1\dots B_N\}$ we are effectively dropping all conditions on the LHS, and condition~(\ref{eq:ReidSepLHS}) reduces to Reid's criterion~\cite{ReidPRA1989,ReidRMP2009} whose violation indicates steering between $A$ and $B$. For other partitions $\Lambda$, a violation of~(\ref{eq:ReidSepLHS}) indicates that LHS models, if they exist, must necessarily be entangled in the partition $\Lambda$.

Again, we can derive a weaker condition if we replace the local conditional variances by their upper bounds without measurement assistance, \ie the variance of the local reduced density matrices. In this case, we recover a modified uncertainty relation whose violation indicates the inseparability of Bob's reduced density matrix~\cite{GessnerPRA2016,GessnerQuantum2017}.
 
\section{State-independent bounds for genuine multipartite entanglement}
The criteria presented so far distinguish between LHS models whose LHS are separable in a specific partition, but they do not exclude convex combinations of entire families of partitions. For the characterization of multipartite entanglement, it is natural to include the convex hull of partitions with similar properties into the same separability class, \eg in terms of the largest entangled subset (also called entanglement depth or $k$-producibility) or the number of separable subsets ($k$-separability)~\cite{GuhneToth,GuhneTothBriegel}. A systematic classification of this kind can be achieved by representing each partition $\Lambda$ by a Young diagram, whose width $w=\max \Lambda:=\max\{|\mathcal{B}_1|,\dots,|\mathcal{B}_{|\Lambda|}|\}$ and height $h=|\Lambda|$ then identify their $w$-producibility and $h$-separability~\cite{SzalayQUANTUM2019}. Combining these quantities, we can introduce classes of $(w,h)$-separable states whose separable partitions have a width that does not exceed $w$ and a height no smaller than $h$~\cite{RenPRL2021}. We can make use of the fact that the metrological sensitivity of $(w,h)$-separable states is limited to derive criteria that test the separability properties of LHS models. Here, we focus on collective rotations of $N$ spin-$1/2$ particles generated by $J_z$, but these results can be easily extended to higher-dimensional systems.

Without assistance from Alice, Bob's ability to estimate a local phase shift is determined by the sensitivity properties of his reduced state $\rho^B$. If $\rho^B$ is $(w,h)$-separable, its sensitivity cannot exceed the bound $F_Q[\rho^B,J_z]\leq w(N-h)+N$~\cite{RenPRL2021}. This bound implies widely used entanglement criteria for $w$-producible states~\cite{TothPRA2012,HyllusPRA2012} when the information provided by $h$ is ignored and demonstrates, in particular, that fully separable systems with $(w,h)=(1,N)$ are limited to a sensitivity at the shot-noise limit $F_Q[\rho^B,J_z]\leq N$ whereas genuine $N$-partite entangled states with $(w,h)=(N,1)$ can in principle reach the Heisenberg limit $F_Q[\rho^B,J_z]\leq N^2$~\cite{GiovannettiPRL2006}.

This paradigm, however, can be broken by assistance from a remote system, Alice. If a $(w,h)$-separable LHS model exists, Bob's assisted sensitivity~(\ref{eq:assistedFQ}) is bounded by 
\begin{align}\label{eq:whsepFQ}
F_Q^{B|A}[\assem,J_z]\leq w(N-h)+N,
\end{align}
which follows using $F_Q^{B|A}[\assem,J_z]\leq \sum_{\lambda}p(\lambda)F_Q[\sigma^B_{\lambda},J_z]$ for LHS models~\cite{YadinNC2021}, together with the separability limit for each conditional state $\sigma^B_{\lambda}$. The condition~(\ref{eq:whsepFQ}) can be violated, allowing Bob to improve his average phase sensitivity beyond the shot-noise limit, even if his reduced state $\rho^B$ is separable. Interestingly, no steering from Alice to Bob is required (in contrast to the scenario that was considered in Ref.~\cite{GessnerEPJQT2019}) and even purely classical correlations between $A$ and $B$ are sufficient for quantum-enhanced assisted metrology. The state-independent bounds for $(w,h)$-entanglement are indicated in gray in Fig.~\ref{fig:exGHZ}.

\section{Assisted entanglement}
Our approach is not limited to applications in quantum metrology. In the following, we will outline how any convex entanglement witness or quantifier can be converted into a witness or quantifier of the assisted entanglement that can be extracted from conditional states if appropriate information about measurements on another system is made available. Consider a convex function $\mathcal{E}(\rho)\leq \sum_kp_k\mathcal{E}(\psi_k)$, where $\rho=\sum_kp_k\rho_k$, with the property $\mathcal{E}(\rho)>0\Rightarrow$ $\rho$ is entangled. We define the corresponding quantum conditional function as
\begin{align}\label{eq:assistedent}
    \mathcal{E}^{B|A}_Q(\assem):=\max_X\sum_ap(a|X)\mathcal{E}(\rho^B_{a|X}).
\end{align}
Convexity implies that $\mathcal{E}^{B|A}_Q(\assem)\leq \sum_{\lambda}p(\lambda)\mathcal{E}(\sigma_{\lambda})$ whenever a LHS model exists. If additionally, the $\sigma_{\lambda}$ are separable, we obtain the bound $\mathcal{E}^{B|A}_Q(\assem)\leq 0$. Hence, we find that
\begin{align}\label{eq:assistedentcrit}
    \mathcal{E}^{B|A}_Q(\assem)>0 \Rightarrow \text{no separable LHS model exists}.
\end{align}
Independently of whether steering is possible or not, $\mathcal{E}^{B|A}_Q(\assem)>0$ reveals that entanglement is present in Bob's conditional states and that it can be made available by suitable measurements on Alice's possibly remote system. Any quantum information protocol that requires entanglement can be converted into an assisted protocol in which Alice communicates her measurement setting and result to Bob. The assistance by Alice may enable Bob to implement a task for which otherwise he would not possess the required resources, \ie when his reduced state is separable, $\mathcal{E}(\rho^B)=0$. The existence of inseparable LHS models further implies that this is possible even in the absence of steering from Alice to Bob.

The metrological criterion for separable LHS~(\ref{eq:whsepFQ}) is a special case of Eq.~(\ref{eq:assistedentcrit}). Other conceivable applications are assisted quantum teleportation protocols, where Bob aims to teleport a state from one of his subsystems to another using the entanglement~\cite{BennettPRL1993} that is made available to him by assistance from Alice, or similarly the implementation of a secure quantum key distribution protocol between subsystems of Bob based on the violation of a Bell's inequality~\cite{EkertPRL1991}. This idea also applies to quantitative witnesses, entanglement measures and other quantifiers that may express with what level of fidelity or security Bob is able to implement the task at hand~\cite{GuhneToth,VedralPRL1997,AcinPRL2007,FriisNATREV2019, MorrisPRX,FadelQuant21}. For example, if $\mathcal{E}$ is an entanglement measure, then $\mathcal{E}^{B|A}_Q(\assem)$ quantifies the average assisted entanglement of Bob.

\section{Conclusions}
We have introduced a classification of LHS models in terms of their entanglement and proposed criteria that are able to put quantitative bounds on the separability properties of LHS. We focused on metrological criteria that are experimentally accessible in a variety of experiments, including cold atoms~\cite{StrobelSCIENCE2014,FadelSplitBEC,KunkelSplitBEC,LangeSplitBEC}, trapped ions~\cite{BohnetSCIENCE2016}, and photons~\cite{Qin2019}. In these experiments, it is often challenging to meet the demanding requirements to observe steering. We have derived a family of weaker bounds whose violation reveals the presence of entanglement in Bob's conditional states. A hierarchy of state-dependent bounds converge to steering criteria in the limit of genuinely multipartite entangled, \ie arbitrary LHS.

The entanglement of LHS has a clear operational interpretation and can be exploited for quantum information tasks. Assisted protocols where Alice carries out a suitable measurement and communicates her result and setting to Bob can unlock hidden entanglement from Bob's system. This is possible even if the two share only classical correlations, and even if Bob's reduced state is separable. This approach may also be extended to other convex properties of interest, such as quantum non-Gaussianity~\cite{TakagiPRA2018,AlbarelliPRA2018} and coherence~\cite{StreltsovRMP2017}. By accessing the substructure of unsteerable states, our results show that even multipartite systems that can be accounted for by LHS models may contain nontrivial quantum correlations, which can be used for quantum information tasks.

\section*{Acknowledgments}
MF was supported by the Swiss National Science Foundation, and by The Branco Weiss Fellowship -- Society in Science, administered by the ETH Z\"{u}rich.
MG acknowledges funding from the LabEx ENS-ICFP: ANR-10-LABX-0010 / ANR-10-IDEX-0001-02 PSL, from MCIN / AEI for the project PID2020-115761RJ-I00, and support of a fellowship from ”la Caixa” Foundation (ID 100010434) and from the European Union’s Horizon 2020 research and innovation programme under the Marie Sk\l{}odowska-Curie grant agreement No 847648, fellowship code LCF/BQ/PI21/11830025.

\clearpage
\newpage

\begin{widetext}


\section{Appendix: Proof of Eq.~(\ref{eq:mainLambda})}\label{si:proof}

Let us assume a $\Lambda$-separable LHS model, \ie the assemblage
\begin{align}\label{eq:sepLHS}
    \assem(a,X) &= \sum_\lambda  p(\lambda)  p(a|X,\lambda) \sigma^{B_1\dots B_N}_\lambda,
\end{align}
where each conditional state is $\Lambda$-separable, namely
\begin{align}
    \sigma^{B_1\dots B_N}_\lambda&=\sum_{\gamma}p_{\lambda}(\gamma)\sigma^{\mathcal{B}_1}_{\lambda,\gamma}\otimes\cdots\otimes \sigma^{\mathcal{B}_{|\Lambda|}}_{\lambda,\gamma} \;.
\end{align}
In the assisted protocol, Bob's average sensitivity is bounded by
\begin{align}\label{eq:qfiLHS}
    \qfi^{B_1\dots B_N \vert A}[\assem,H]  &\leq \sum_{\lambda}p(\lambda)\qfi[\sigma^{B_1\dots B_N}_\lambda, H]\notag\\
    &\leq 4\sum_{\lambda}p(\lambda)\sum_{k=1}^{|\Lambda|} \var[\sigma^{\mathcal{B}_k}_{\lambda}, H^{\mathcal{B}_k}]
\end{align}
where
\begin{align}
    \sigma^{\mathcal{B}_k}_{\lambda}=\sum_{\gamma}p_{\lambda}(\gamma)\sigma^{\mathcal{B}_k}_{\lambda,\gamma},
\end{align}
is the reduced density matrix on subsystem $\mathcal{B}_k$ for the state $\sigma^{B_1\dots B_N}_{\lambda}$, and
\begin{align}
    \sigma^{\mathcal{B}_k}=\sum_{\lambda}p(\lambda)\sum_{\gamma}p_{\lambda}(\gamma)\sigma^{\mathcal{B}_k}_{\lambda,\gamma},
\end{align}
is the corresponding reduced density matrix for the full state $\sigma^{B}$. In the first step, we used the convexity of the QFI, which leads to an upper bound on the QCFI for LHS models~\cite{YadinNC2021}. In the second step, we made use of an upper bound on the QFI for $\Lambda$-separable states in terms of local variances~\cite{GessnerPRA2016}.
Finally, we used the concavity of the variance to obtain a lower bound on the QCV in the presence of LHS descriptions~\cite{YadinNC2021}, for each local subsystem.

Finally, consider the measurement of the local generator $H^{\mathcal{B}_k}$ in subsystem $\mathcal{B}_k$, assisted by Alice's communication about her result $a$ and setting $X$. The existence of the LHS model~(\ref{eq:sepLHS}) between Alice and all of Bob's subsystems implies the assemblage $\assem(a,X)=\sum_{\lambda}p(\lambda)p(a|X,\lambda)\sigma^{\mathcal{B}_k}_{\lambda}$ for each individual subsystem $\mathcal{B}_k$ after tracing out the remaining subsystems. Making use of the lower bound on the QCV for LHS~\cite{YadinNC2021,ReidRMP2009} we obtain
\begin{align}\label{eq:varLHS}
	\var_Q^{\mathcal{B}_k|A}[\assem,H^{\mathcal{B}_k}] 	&\geq \sum_\lambda p(\lambda) \var[\sigma^{\mathcal{B}_k}_\lambda, H^{\mathcal{B}_k}].
\end{align}
Combining Eqs.~(\ref{eq:qfiLHS}) and~(\ref{eq:varLHS}) we obtain
\begin{align}
    \qfi^{B_1\dots B_N \vert A}[\assem,H]   &\leq 4\sum_{k=1}^{|\Lambda|}\var_Q^{\mathcal{B}_k|A}[\assem,H^{\mathcal{B}_k}],
\end{align}
which is the result presented in Eq.~(\ref{eq:mainLambda}) of the main text.

\clearpage
\newpage

\end{widetext}


\begin{thebibliography}{}

\bibitem{NielsenChuang} M. A. Nielsen and I. L. Chuang, \textit{Quantum Computation and Quantum Information} (Cambridge University Press, New York, NY, 2000).

\bibitem{HorodeckiRMP} R. Horodecki, P. Horodecki, M. Horodecki, and K. Horodecki, Quantum entanglement, \href{https://doi.org/10.1103/RevModPhys.81.865}{Rev. Mod. Phys. \textbf{81}, 865 (2009)}.

\bibitem{ReidRMP2009} M. D. Reid, P. D. Drummond, W. P. Bowen, E. G. Cavalcanti, P. K. Lam, H. A. Bachor, U. L. Andersen, and G. Leuchs, Colloquium: The Einstein-Podolsky-Rosen paradox: From concepts to applications, \href{https://doi.org/10.1103/RevModPhys.81.1727}{Rev. Mod. Phys. \textbf{81}, 1727 (2009)}.

\bibitem{BrunnerRMP} N. Brunner,  D. Cavalcanti,  S. Pironio,  V. Scarani,  and S. Wehner, Bell Nonlocality, \href{https://doi.org/10.1103/RevModPhys.86.419}{Rev. Mod. Phys. \textbf{86}, 419 (2014)}

\bibitem{CavalcantiREPPROGPHYS17} D. Cavalcanti and P. Skrzypczyk, Quantum steering: a review with focus on semidefinite programming, \href{https://doi.org/10.1088/1361-6633/80/2/024001}{Rep. Prog. Phys. \textbf{80}, 024001 (2017)}.

\bibitem{FriisNATREV2019} N. Friis, G. Vitagliano, M. Malik, and M. Huber,
Entanglement certification from theory to experiment, \href{https://doi.org/10.1038/s42254-018-0003-5}%
{Nat. Rev. Phys. \textbf{1}, 72 (2019)}.

\bibitem{GuehneRMP2020} R. Uola, A. C. S. Costa, H. C. Nguyen, and O. G\"{u}hne, Quantum steering, \href{https://doi.org/10.1103/RevModPhys.92.015001}{Rev. Mod. Phys. \textbf{92}, 015001 (2020)}.

\bibitem{DurPRA2000} W. D\"ur, G. Vidal, and J. I. Cirac, Three qubits can be entangled in two inequivalent ways, \href{https://doi.org/10.1103/PhysRevA.62.062314}{Phys. Rev. A \textbf{62}, 062314 (2000)}.

\bibitem{GuhneToth} O. G\"uhne and G. T\'oth, Entanglement Detection, \href{https://doi.org/10.1016/j.physrep.2009.02.004}{Phys. Rep. \textbf{474}, 1 (2009)}.


\bibitem{LeviPRL2013}
F. Levi and F. Mintert, Hierarchies of Multipartite Entanglement, \href{https://doi.org/10.1103/PhysRevLett.110.150402}{Phys. Rev. Lett. \textbf{110}, 150402 (2013)}.

\bibitem{WalterArXiv}
M. Walter, D. Gross, J. Eisert, Multi-partite entanglement, \href{https://arxiv.org/abs/1612.02437}{arXiv:1612.02437}.

\bibitem{SzalayQUANTUM2019} S. Szalay, k-stretchability of entanglement,
and the duality of k-separability and k-producibility, \href{https://doi.org/10.22331/q-2019-12-02-204}%
{Quantum \textbf{3}, 204 (2019)}.

\bibitem{Bell1964} J.-S. Bell, Physics (Long Island City, N.Y.) \textbf{1}, 195 (1964).

\bibitem{WisemanPRL2007} H. M. Wiseman, S. J. Jones, and A. C. Doherty, Steering, Entanglement, Nonlocality, and the Einstein-Podolsky-Rosen Paradox, \href{https://doi.org/10.1103/PhysRevLett.98.140402}{Phys. Rev. Lett. \textbf{98}, 140402 (2007)}.

\bibitem{GallegoPRX2015} 
R. Gallego and L. Aolita, Resource Theory of Steering, \href{https://doi.org/10.1103/PhysRevX.5.041008}{Phys. Rev. X \textbf{5}, 041008 (2015)}.

\bibitem{EPR1935} A. Einstein, B. Podolsky, and N. Rosen, Can Quantum-Mechanical Description of Physical Reality Be Considered Complete?, \href{https://doi.org/10.1103/PhysRev.47.777}{Phys. Rev. \textbf{47}, 777 (1935)}.

\bibitem{ReidPRA1989} M. D. Reid, Demonstration of the Einstein-Podolsky-Rosen paradox using nondegenerate parametric amplification, \href{https://doi.org/10.1103/PhysRevA.40.913}{Phys. Rev. A \textbf{40}, 913 (1989)}.

\bibitem{YadinNC2021} B. Yadin, M. Fadel, and M. Gessner, Metrological complementarity reveals the Einstein-Podolsky-Rosen paradox, \href{https://doi.org/10.1038/s41467-021-22353-3 }{Nat. Commun. \textbf{12}, 2401 (2021)}.

\bibitem{GuoArXiv} J. Guo, F.-X. Sun, D. Zhu, M. Gessner, Q. He, M. Fadel, Detecting Einstein-Podolsky-Rosen steering in non-Gaussian spin states from conditional spin-squeezing parameters, \href{https://arxiv.org/abs/2106.13106}{arXiv:2106.13106}.

\bibitem{WalbornPRL11}
S. P. Walborn, A. Salles, R. M. Gomes, F. Toscano, and P. H. Souto Ribeiro, Revealing Hidden Einstein-Podolsky-Rosen Nonlocality, \href{https://doi.org/10.1103/PhysRevLett.106.130402}{Phys. Rev. Lett. \textbf{106}, 130402 (2013)}.

\bibitem{WalbornPRA13} J. Schneeloch, C. J. Broadbent, S. P. Walborn, E. G. Cavalcanti and J. C. Howell, Einstein-Podolsky-Rosen steering inequalities from entropic uncertainty relations, \href{https://doi.org/10.1103/PhysRevA.87.062103}{Phys. Rev. A \textbf{87}, 062103 (2013)}.

\bibitem{CostaEntropy2018}
A. C. S. Costa, R. Uola, O. G\"uhne, Entropic Steering Criteria: Applications to Bipartite and Tripartite Systems, \href{https://doi.org/10.3390/e20100763}{Entropy \textbf{20}, 763 (2018)}.

\bibitem{RiccardiPRA2018}
A. Riccardi, C. Macchiavello, and L. Maccone, Multipartite steering inequalities based on entropic uncertainty relations, \href{https://doi.org/10.1103/PhysRevA.97.052307 }{Phys. Rev. A \textbf{97}, 052307 (2018)}.

\bibitem{HePRL2013}
Q. Y. He and M. D. Reid, Genuine Multipartite Einstein-Podolsky-Rosen Steering, \href{https://doi.org/10.1103/PhysRevLett.111.250403}{Phys. Rev. Lett. \textbf{111}, 250403 (2013)}.

\bibitem{ReidPRA2013}
M. D. Reid, Monogamy inequalities for the Einstein-Podolsky-Rosen paradox and quantum steering, \href{https://doi.org/10.1103/PhysRevA.88.062108}{Phys. Rev. A \textbf{88}, 062108 (2013)}.

\bibitem{TehPRA2014}
R. Y. Teh and M. D. Reid,
Criteria for genuine N-partite continuous-variable entanglement and Einstein-Podolsky-Rosen steering,
\href{https://doi.org/10.1103/PhysRevA.90.062337}{Phys. Rev. A \textbf{90}, 062337 (2014)}.

\bibitem{ArmstrongNatPhys2015}
S. Armstrong, M. Wang, R. Y. Teh, Q. Gong, Q. He, J. Janousek, H.-A. Bachor, M. D. Reid, P. K. Lam, Multipartite Einstein–Podolsky–Rosen steering and genuine tripartite entanglement with optical networks, \href{https://doi.org/10.1038/nphys3202}{Nature Phys. \textbf{11}, 167 (2015)}.

\bibitem{CavalcantiNatCommun2015} D. Cavalcanti, P. Skrzypczyk, G. H. Aguilar, R. V. Nery, P.H. Souto Ribeiro, S. P. Walborn, Detection of entanglement in asymmetric quantum networks and multipartite quantum steering, \href{https://doi.org/10.1038/ncomms8941}{Nat. Commun. \textbf{6}, 7941 (2015)}.

\bibitem{XiangPRA2019}
Y. Xiang, X. Su, L. Mi\v{s}ta, Jr., G. Adesso, and Q. He, Multipartite Einstein-Podolsky-Rosen steering sharing with separable states, \href{https://doi.org/10.1103/PhysRevA.99.010104}{Phys. Rev. A \textbf{99}, 010104(R) (2019)}.

\bibitem{BraunsteinPRL1994} S. L. Braunstein and C. M. Caves, Statistical
Distance and the Geometry of Quantum States, \href{https://doi.org/10.1103/PhysRevLett.72.3439}%
{Phys. Rev. Lett. \textbf{72}, 3439 (1994)}.

\bibitem{GiovannettiNatPhoton2011} V. Giovannetti, S. Lloyd, and L. Maccone, Advances in quantum metrology, \href{https://doi.org/10.1038/nphoton.2011.35}{Nature Phot. \textbf{5}, 222 (2011)}.

\bibitem{ApellanizJPA2014} G. T\'{o}th and I. Apellaniz, Quantum metrology
from a quantum information science perspective, \href{https://doi.org/10.1088/1751-8113/47/42/424006}%
{J. Phys. A \textbf{47}, 424006 (2014)}.

\bibitem{Varenna} L. Pezz\`e and A. Smerzi, \textit{Quantum theory of phase
estimation}, in \textit{Atom Interferometry, Proceedings of the
International School of Physics ``Enrico Fermi'', Course 188, Varenna},
edited by G. M. Tino and M. A. Kasevich (IOS Press, Amsterdam) p. 691 (2014).

\bibitem{Supp} See the Supplemental Material for a proof of Eq.~(\ref{eq:mainLambda}).

\bibitem{GessnerPRA2016} M. Gessner, L. Pezz\`{e}, A. Smerzi, Efficient entanglement criteria for discrete, continuous, and hybrid variables, \href{https://doi.org/10.1103/PhysRevA.94.020101}{Phys. Rev. A \textbf{94}, 020101(R) (2016)}.


\bibitem{Cohen98} O. Cohen, Unlocking Hidden Entanglement with Classical Information, \href{https://doi.org/10.1103/PhysRevLett.80.2493}{Phys. Rev. Lett. \textbf{80}, 2493 (1998)}.

\bibitem{Cohen99} O. Cohen, Unlocking Hidden Entanglement with Classical Information, \href{https://doi.org/10.1103/PhysRevA.60.80}{Phys. Rev. A \textbf{60}, 80 (1999)}.


\bibitem{GHZ} D. M. Greenberger, M. A. Horne, and A. Zeilinger, Going Beyond Bell's Theorem, in ``Bell's Theorem, Quantum Theory, and Conceptions of the Universe'', M. Kafatos (Ed.), Kluwer, Dordrecht, pp. 69 (1989).

\bibitem{HofmannPRA2003} H. F. Hofmann and S. Takeuchi, Violation of local uncertainty relations as a signature of entanglement, \href{https://doi.org/10.1103/PhysRevA.68.032103}{Phys. Rev. A \textbf{68}, 6 (2003)}.

\bibitem{FadelPRA2020} M. Fadel and M. Gessner, Relating spin squeezing to multipartite entanglement criteria for particles and modes, \href{https://doi.org/10.1103/PhysRevA.102.012412}{Phys. Rev. A \textbf{102}, 012412 (2020)}.

\bibitem{GessnerQuantum2017}
M. Gessner, L. Pezz\`e, and A. Smerzi,
Entanglement and squeezing in continuous-variable systems,
\href{https://doi.org/10.22331/q-2017-07-14-17}{Quantum {\bf 1}, 17 (2017)}.

\bibitem{GuhneTothBriegel} O. G\"uhne, G. T\'oth anf H. Briegel, Multipartite entanglement in spin chains, \href{https://doi.org/10.1088/1367-2630/7/1/229}{New J. Phys. \textbf{7}, 229 (2005)}.

\bibitem{RenPRL2021} Z. Ren, W. Li, A. Smerzi, and M. Gessner, Metrological Detection of Multipartite Entanglement from Young Diagrams, \href{https://doi.org/10.1103/PhysRevLett.126.080502}{Phys. Rev. Lett. \textbf{126}, 080502 (2021)}.

\bibitem{HyllusPRA2012} P. Hyllus, W. Laskowski, R. Krischek, C.
Schwemmer, W. Wieczorek, H. Weinfurter, L. Pezz\`e, and A. Smerzi, Fisher
information and multiparticle entanglement, \href{https://doi.org/10.1103/PhysRevA.85.022321}{Phys. Rev. A \textbf{85}, 022321 (2012)}.

\bibitem{TothPRA2012} G. T\'{o}th, Multipartite entanglement and high-precision metrology, \href{https://doi.org/10.1103/PhysRevA.85.022322}{Phys. Rev. A \textbf{85}, 022322 (2012)}.

\bibitem{GiovannettiPRL2006}
V. Giovannetti, S. Lloyd, and L. Maccone, Quantum metrology,
\href{https://doi.org/10.1103/PhysRevLett.96.010401}{Phys. Rev. Lett. \textbf{96}, 010401 (2006)}.

\bibitem{GessnerEPJQT2019} M. Gessner and A. Smerzi, Encrypted quantum correlations: Delayed choice of quantum statistics and other applications, \href{https://doi.org/10.1140/epjqt/s40507-019-0074-y}{EPJ Quantum Technol. \textbf{6}, 4 (2019).}

\bibitem{BennettPRL1993} C. H. Bennett, G. Brassard, C. Cr\'{e}peau, R. Jozsa, A. Peres, and W. K. Wootters, \href{https://doi.org/10.1103/PhysRevLett.70.1895}{Phys. Rev. Lett. \textbf{70}, 1895 (1993)}.

\bibitem{EkertPRL1991} A. K. Ekert, Quantum cryptography based on Bell’s theorem, \href{https://doi.org/10.1103/PhysRevLett.67.661}{Phys. Rev. Lett. \textbf{67}, 661 (1991)}.

\bibitem{VedralPRL1997}
V. Vedral, M. B. Plenio, M. A. Rippin, and P. L. Knight, Quantifying Entanglement, \href{https://doi.org/10.1103/PhysRevLett.78.2275}{Phys. Rev. Lett. \textbf{78}, 2275 (1997)}.

\bibitem{AcinPRL2007}
A. Ac\'{i}n, N. Brunner, N. Gisin, S. Massar, S. Pironio, and V. Scarani, Device-Independent Security of Quantum Cryptography against Collective Attacks, \href{https://doi.org/10.1103/PhysRevLett.98.230501}{Phys. Rev. Lett. \textbf{98}, 230501 (2007)}.

\bibitem{MorrisPRX}
B. Morris, B. Yadin, M. Fadel, T. Zibold, P. Treutlein and G. Adesso, Entanglement between identical particles is a useful and consistent resource, \href{https://doi.org/10.1103/PhysRevX.10.041012}{Phys. Rev. X \textbf{10}, 041012 (2020)}.

\bibitem{FadelQuant21}
M. Fadel, A. Usui, M. Huber, N. Friis and G. Vitagliano, Entanglement Quantification in Atomic Ensembles, \href{https://doi.org/10.1103/PhysRevLett.127.010401}{Phys. Rev. Lett. \textbf{127}, 010401 (2021)}.

\bibitem{StrobelSCIENCE2014}
H. Strobel, W. Muessel, D. Linnemann, T. Zibold, D. B. Hume, L. Pezz\`e, A. Smerzi, and M. K. Oberthaler, 
Fisher information and entanglement of non-Gaussian spin states, 
\href{https://doi.org/10.1126/science.1250147 }{Science {\bf 345}, 424 (2014)}.

\bibitem{FadelSplitBEC}
M. Fadel, T. Zibold, B. D\'{e}camps and P. Treutlein, Spatial entanglement patterns and Einstein-Podolsky-Rosen steering in Bose-Einstein condensates, \href{https://doi.org/10.1126/science.aao1850}{Science \textbf{360}, 409 (2018)}.

\bibitem{KunkelSplitBEC}
P. Kunkel, M. Pr\"ufer, H. Strobel, D. Linnemann, A. Fr\"olian, T. Gasenzer, M. G\"arttner and M. K. Oberthaler, Spatially distributed multipartite entanglement enables EPR steering of atomic clouds, \href{https://doi.org/10.1126/science.aao2254}{Science \textbf{360}, 413 (2018)}.

\bibitem{LangeSplitBEC}
K. Lange, J. Peise, B. L\"ucke, I. Kruse, G. Vitagliano, I. Apellaniz, M. Kleinmann, G. T\'oth and C. Klempt, Entanglement between two spatially separated atomic modes, \href{https://doi.org/10.1126/science.aao2035}{Science \textbf{360}, 416 (2018)}.

\bibitem{BohnetSCIENCE2016} J. G. Bohnet, B. C. Sawyer, J. W. Britton, M. L. Wall, A. M. Rey, M. Foss-Feig, J. J. Bollinger, Quantum spin dynamics and entanglement generation with hundreds of trapped ions, \href{https://doi.org/10.1126/science.aad9958 }{Science \textbf{352}, 1297 (2016)}.

\bibitem{Qin2019} Z. Qin, M. Gessner, Z. Ren, X. Deng, D. Han, W. Li, X. Su, A. Smerzi, and K. Peng, Characterizing the multipartite continuous-variable entanglement structure from squeezing coefficients and the Fisher information, \href{https://doi.org/10.1038/s41534-018-0119-6}{npj Quant. Inf. \textbf{5}, 3 (2019)}.

\bibitem{TakagiPRA2018}
R. Takagi and Q. Zhuang, Convex resource theory of non-Gaussianity, \href{https://doi.org/10.1103/PhysRevA.97.062337}{Phys. Rev. A \textbf{97}, 062337 (2018)}.

\bibitem{AlbarelliPRA2018}
F. Albarelli, M. G. Genoni, M. G. A. Paris, and A. Ferraro, Resource theory of quantum non-Gaussianity and Wigner negativity, \href{https://doi.org/10.1103/PhysRevA.98.052350}{Phys. Rev. A \textbf{98}, 052350 (2018)}.

\bibitem{StreltsovRMP2017} 
A. Streltsov, G. Adesso, and M. B. Plenio, Colloquium: Quantum coherence as a resource, \href{https://doi.org/10.1103/RevModPhys.89.041003}{Rev. Mod. Phys. \textbf{89}, 041003 (2017)}.




\end{thebibliography}
\end{document}